# Whisphone: Whispering Input Earbuds

FUKUMOTO Masaaki （福本 雅朗）*

Abstract. Whisphone is a novel earbud device designed for speech input via whispering. Utilizing canal-type earbuds with a unique microphone placement at the tip of the earplug, it effectively captures whispered voices radiated in the ear canal through bone conduction. This design can boost whispered voice volume with ear canal occlusion effect while simultaneously blocking external noise by sealing the ear hole. By incorporating Active Noise Canceling (ANC), Whisphone can effectively detect subtle whispers, even in noisy environments of up to 80dB(A). Its compact and comfortable design ensures discreet wearability, allowing users to interact with AI assistants hands-free without disturbing others in various daily situations such as offices, homes, or urban public spaces.

## 1 Introduction

The evolution of Large Language Models (LLMs) and generative AI technologies has transformed the way we interact with computers, elevating them from mere "operating machines" to "conversational partners". Speech-based interactions are expected to play a prominent role in this shift; however, a critical challenge persists: sound leakage from audio input. Not only does it disturb nearby individuals but also raises concerns about privacy and potential information leaks. Imagining a future where people loudly speak into their devices in confined spaces like offices or on bustling streets is far from ideal for maintaining peace and confidentiality. To address this issue and ensure privacy in various environments, there is a growing need to develop an alternative speech input method that eliminates the leakage of uttered voices.

### 1.1 Silent Voice Input

Silent voice input methods can be categorized into two main types: fully silent (producing no sound) and those that emit faint sounds that are imperceptible to others. The first category, completely silent input methods, typically involves the use of sensors to capture movements associated with articulation in and around the mouth cavity. Sottovoce [13], for example, employs an ultrasonic sensor under the chin to detect jaw, lip, and tongue movements, which are then converted into speech. Other methods may utilize accelerometers [18] or EMG (Electromyogram) electrodes [28] attached to the cheeks or jaw. Some approaches also include the use of accelerometers [6] or strain gauges [9] integrated into masks, or even reflected sound waves radiated into the ear canal through earbuds [3]. There are also technologies that detect EEG (Electroencephalogram) signals from electrodes placed on the head when a user imagines speaking without any mouth movement [4].

Regardless of the sensing mechanism employed, specialized recognition systems are necessary to convert sensor signals into understandable text or audible speech. These mechanisms often require additional learning tailored to individual users and may be limited in their ability to recognize specific phonemes or words. Real-time processing capabilities also vary among existing approaches.

### 1.2 Whisper-based Speech Input

In contrast, methods that detect faint sounds, such as whispers, offer a distinct advantage over entirely silent approaches because they capture signals similar to those of normal speech. This similarity makes it relatively easier to design recognition mechanisms, often resulting in higher performance in terms of accuracy and speed.

Pioneering studies in this field is NAM (Non-Audible Murmur) [12], which uses a "flesh conduction microphone" that comes into contact with the Whisphone: Whispering Input Earbudsback lower part of the ear to capture faint murmurs for speech input while minimizing sound leakage. SilentVoice [11] introduces an innovative technique where speaking while *inhaling* allows the microphone to be positioned extremely close without generating pop noise, capturing ultra-soft voices below 40dB(A). WhisperMask [5], featuring a large flat film-shaped ECM (Electret Condenser Microphone) inside a mask, takes advantage of its superior close-talk performance to clearly detect whispered speech even in noisy environments. While most approaches recognize captured whispers as text, WESPER [17] convert them into regular audible speech in real time, enabling conversations not only with computers but also between individuals.

There are some speech recognition mechanisms capable of understanding whispered voices; examples include Google Voice Input, OpenAI's Whisper and Amazon Alexa's "Whisper Mode". These advancements allow users to input commands or interact without disturbing their surroundings. Additionally, the ability to distinguish between normal speech and whispers enables the use of whispers as commands [16], adding another layer of versatility to voice input.

However, effectively capturing quiet whispers often requires speaking close to devices like smartphones or wearing headset microphones near the mouth. Earbud-type headsets may employ beamforming techniques to amplify speech, but they can still face challenges in noisy environments or busy streets. The convenience and aesthetics of wearing headset microphones throughout the day are also factors that may hinder their widespread adoption by the general public, especially during activities like eating.

1.3 Bone Conduction Microphone

Most of the whisper input mechanisms mentioned above detect airborne sounds radiated into the surrounding atmosphere during speech. In contrast, bone conduction methods (including NAM, which falls within this category) capture vibrations that propagate through the skull and soft tissues to a sensor placed on the skin surface. Typical bone conduction pickup devices are laryngeal microphones [22], which contacts against the throat to detect vocal cord vibrations. Other techniques involve detecting vibrations from the walls of the external ear canal [21], using small ECM microphones placed inside the ear canal [20] to capture radiated sounds, and utilizing MEMS vibration sensors mounted on earbuds to sense vibrations around the earhole [19]. Additionally, some commercially available noise-canceling earbuds improve speech intelligibility by incorporating accelerometers within their housings to detect speech vibrations [7].

While bone conduction methods offer robustness against external noise, they suffer from significant attenuation of high-frequency components, resulting in muffled-sounding voices. Consequently, their use has typically been limited to specific scenarios, such as high-noise environments. Except for NAM, most products target regular speech input rather than capturing faint whispers effectively.

1.4 Capturing Whispered Voice via Bone Conduction

The commonly known pathway for bone conduction hearing is the sound perception through vibrations traveling via the skull and soft tissues to reach the ossicles or cochlea directly. However, there is an additional route called "(external) ear canal radiation", where these acoustic vibrations cause the wall of the external ear canal to vibrate, radiating sound within the canal. This radiated sound then reaches the eardrum, resulting in hearing just like regular airborne sounds. It has been suggested that this latter pathway significantly contributes to bone conduction hearing in individuals with normal hearing [25].

Additionally, a phenomenon known as the "(external) ear canal occlusion effect" occurs when the external ear canal opening (= ear hole) is blocked by fingers or earplugs. This blocking results in an amplification of low-frequency components below 1kHz by 5-20dB during bone conduction listening [8]. The amplification occurs because the acoustic energy radiated into the ear canal through the latter pathway is prevented from escaping through the ear canal opening.

Since this effect also applies to our perception of own vocal feedback, it is assumed that whispered voices and other faint utterances can be recorded in an amplified state by placing a microphone inside the blocked ear canal. This setup offers the additional advantage of blocking external noise

Whisphone: Whispering Input Earbuds

from reaching the microphone due to the physical seal created within the ear canal.

However, whisper volumes are typically around 40dB(A), which is significantly lower than regular speech at approximately 60dB(A). Even with the assumed amplification of bone conduction components and reduction of external noise due to ear canal occlusion effect, achieving a favorable S/N (Signal-to-Noise) ratio in high-noise environments remains challenging.

On the other hand, with the widespread adoption of TWS (True Wireless Stereo) earbuds, advancements in ANC (Active Noise Canceling) technology have been remarkable. Many canal-type earbuds claim noise suppression performance exceeding 45dB according to their specifications [2]. Considering that wearing canal-type earbuds is analogous to blocking the external ear canal opening, combining them with ANC functionality should provide similar noise cancellation effects for microphones placed inside the ear canal.

In this paper, we introduce Whisphone, a full-time wearable whisper input device. It records bone conducted whispered speech radiated from the walls of the external ear canal through a microphone located at the tip of an earplug-type canal earbud. By combining this setup with ANC (Active Noise Canceling) technology, it is possible to capture faint whispering sounds without disturbing others even in noisy environments. In the following chapters, we will present experimental results and explanations regarding the amplification of bone conduction speech due to external ear canal occlusion effect and the combined noise reduction of earplugs and ANC. We will then show that whisper recognition is feasible even in noisy conditions by integrating our setup with a generic speech recognition mechanism, eliminating the need for additional learning or personalized adaptation. Finally, we will provide implementation examples, discuss current challenges and proposed solutions, explore future prospects, and conclude our findings.

## 2 Whisphone

### 2.1 Configuration

Whisphone's configuration is illustrated in Figure 1. By placing a microphone at the tip of an earplug-type canal earbud with ANC functionality, Whisphone captures bone conduction speech radiated within the external ear canal during whispered utterances. The use of canal earbuds effectively seals the opening of the earhole, providing noise reduction and amplification of the audio signal due to the external ear canal occlusion effect. With the additional noise suppression from ANC, whisper input becomes feasible even in noisy environments.

### 2.2 Sound Pickup Performance

To evaluate Whisphone's recording performance, we installed a small MEMS (Micro Electro Mechanical Systems) microphone [10] at the tip of canal-type earbuds with ANC functionality [2]. The earbuds were connected to a PC via a USB audio interface [1], we adjusted the low-cut filter setting on the interface to 235Hz. During the experiment, an adult male participant wore the earbuds while whispering at a volume of 40dB(A) from a distance of one meter. For simulating background noise, recordings from a train platform were used, calibrated to create external noise levels of either 60dB(A) or 80dB(A) at the ear position. The background noise in the experimental environment was measured at 33dB(A), referred to as the "quiet environment". The ANC function on the earbuds set to its maximum strength in "Quiet" mode.

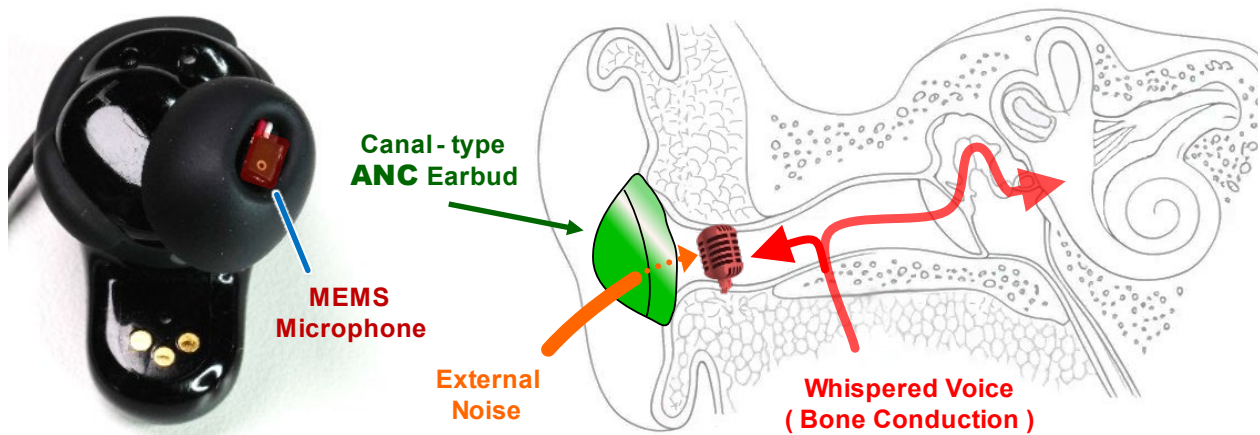

Figure 1. Whisphone's design: A microphone installed at the tip of a canal-type ANC earbud captures whispered speech radiated within the ear canal through bone conduction.

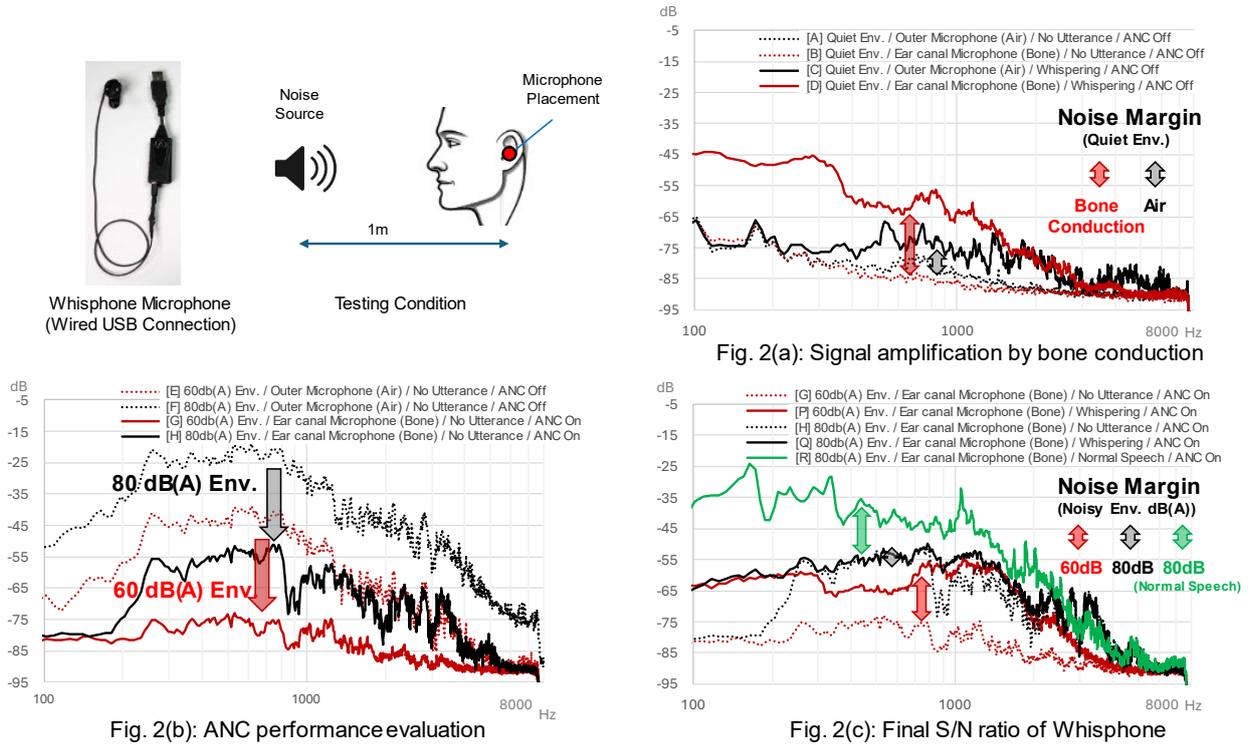

Figure 2. Recording performance of Whisphone: (a) Ear canal occlusion effect amplifies whispered voices by about 10 dB; (b) ANC reduces external noise levels by approximately 30 dB; (c) The overall S/N ratio improvement reaches up to 40dB, enabling whisper input even in noisy environments.

The experiment involved a single subject and utilized the utterance "Konnichiwa, watashi no koe ga kikoe masuka?" ("Hello, can you hear my voice?" in Japanese). We manipulated various conditions by changing external noise levels (quiet environment / 60dB(A) / 80dB(A)), microphone placement (inside the ear canal / outside near the earhole), the presence or absence of utterances, and ANC functionality being activated or deactivated during recordings.

Figure 2 shows the results. Figure 2(a) confirms volume amplification due to ear canal occlusion effect with an ear canal microphone. Compared to the background noise in a silent environment [A], recording air-conducted whispered speech with a same MEMS microphone unit placed outside near the earhole similar as standard earbuds [C], offers only a limited noise margin of about 10dB, making faint whisper acquisition challenging. In contrast, by placing the microphone inside the external ear canal and blocking the opening with an earplug [D] results in a 20dB noise margin for frequencies below 1.5kHz when using the same whisper input. This enhancement is attributed to the reduction of external noise levels due to physical blockage by the earplug [B] and the amplification of bone conduction components radiated into the ear canal through the ear canal occlusion effect. On the other hand, bone-conducted sound pickup shows a rapid decrease in frequencies above 2kHz, with almost no recorded sound above 4.5kHz. There is an another increase in components below 500Hz primarily due to deformation of the external ear canal during speech production [D], which we have not actively utilized in our current implementation (further details will be discussed later).

Figure 2(b) demonstrates external noise reduc-

Whisphone: Whispering Input Earbuds

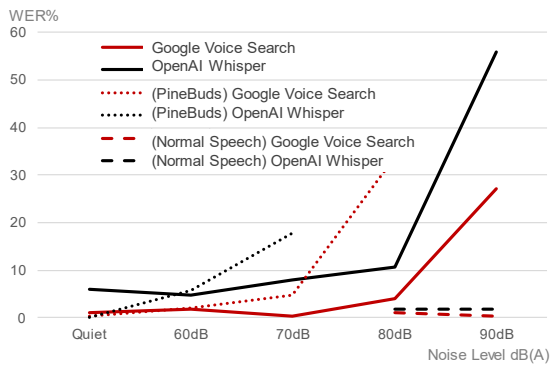

Figure 3. Speech recognition performance of Whisphone

## 2.3 Speech Recognition Performance

Next, we conducted a simple evaluation of Whisphone's speech recognition performance using the same equipment and noise conditions as in the previous section. For this test, we used two widely adopted speech recognition mechanisms: Google's "Voice Search" function on its search page (referred to as "Google Voice Search") and OpenAI's "Whisper" (large-v3 model). No additional learning, personalized adaptation, or enrollment was employed for either system. We selected 15 commonly used Japanese phrases relevant to voice assistant or generative AI control tasks, such as "Ashita no yotei ha nani ga ari masuka?" ("What's my schedule for tomorrow?" in Japanese), "Tokyo eki ha dou ike ba ii desuka?" ("How do I get to Tokyo Station?" in Japanese), and "Atarashii idea wo dashite kudasai" ("Show me some new ideas." in Japanese). Each phrases were whispered three times, and WER (Word Error Rate) values were calculated. The background noise level in the experimental environment was measured at 33dB(A). The experiment was conducted on a single subject, an adult male.

Figure 3 shows the results: up to external noise levels of approximately 80dB(A), similar to the typical noise inside a train carriage, WER values were mostly below 10%, indicating accurate input with faint whispered speech. However, when the noise level increased to 90dB(A), resembling the environment within a subway carriage, WER deteriorated significantly, and VAD (Voice Activity Detection) failures became frequent, rendering whisper recognition impractical in such conditions.

Based on these results, we concluded that Whisphone-based speech recognition is feasible without requiring any additional learning or personalized adaptation when external noise levels are up to 80dB(A). Furthermore, even in environments with external noise exceeding 80dB(A), the use of regular speech volumes continued to yield low WERs for successful voice recognition (as indicated by the dashed lines in Figure 3).

tion through ANC functionality. By combining earplugs with ANC for external noise levels of both 60dB(A) and 80dB(A), approximately 30dB of noise reduction is achieved ([E]vs.[G], [F]vs.[H]).

Figure 2(c) highlights Whisphone's recording performance by combining the ear canal microphone with ANC functionality. In a noisy environment of 60 dB(A), Whisphone successfully captures whispered speech while maintaining a noise margin of up to 20dB ([P]vs.[G]). Considering that whisper volumes are around 40dB(A), Whisphone offers an overall S/N ratio improvement of up to 40dB compared to conventional air-conducted microphones (from -20dB to +20dB). This gain comprises a +10dB boost in signal components due to the ear canal occlusion effect and -30dB reduction in noise components through ANC functionality. Even when faced with external noise levels reaching 80dB(A), Whisphone still achieves a similar S/N ratio improvement of approximately 40dB ([Q]vs.[H]), however the noise margin becomes diminishes, indicating that 80dB(A) is Whisphone's practical limit for accurate whisper recognition tasks.

When the external noise level exceeds this threshold, recognizing whispered speech becomes more challenging. However, using regular speech that has a volume of around 60dB(A) provides an additional 20dB noise margin. This enables effective speech acquisition even in extremely loud environments exceeding 90dB(A) ([R]vs.[H]). It is worth mentioning that when external noise levels reach or exceed 80dB(A), regular speech at a volume of 60dB(A) becomes nearly imperceptible to individuals nearby, ensuring a discreet communication experience with Whisphone.

## 3 Implementation

Next, we present implementation examples of Whisphone. As shown in Chapter 2, the key to en-hancing Whisphone's noise suppression performance lies in ANC functionality. Fortunately, recent advancements have led to the availability of TWS earbuds with rated noise suppression capabilities exceeding 45dB, combining them with miniature microphones allows for simple implementations. Figure 4(a) illustrates an example where we integrated a small MEMS microphone [10] into the tip of the earplug of left earbud of commercially available TWS earbuds [2]. The signal line from the microphone capsule is routed outside using fine cables that maintain the airtightness of the external ear canal, and is connected to a USB audio interface [1] for functioning as a conventional wired microphone. By enabling only the noise cancellation function in offline mode on these TWS earbuds, we can effectively capture noise-suppressed bone conduction signals.

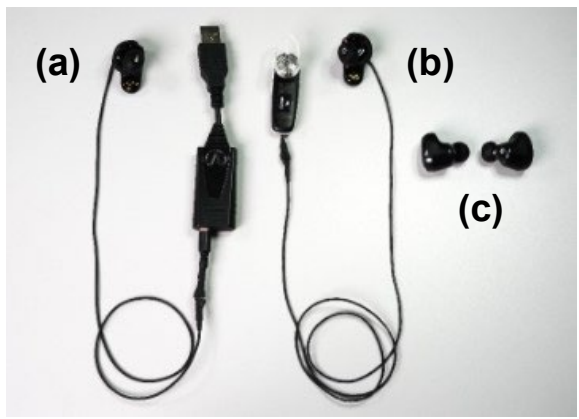

Figure 4. Implementation examples of Whisphone: (a) Wired USB microphone type; (b) Bluetooth monaural headset type; (c) TWS (True Wireless Stereo) earbuds type.

Figure 4(b) shows a modification where we removed the ECM capsule from a commercially available Bluetooth headset [15] and connected it to the small microphone shown in Figure 4(a). With this setup, only the left earbud operates offline with ANC enabled, while the right side functions as a typical monaural Bluetooth headset.

Figure 4(c) presents a fully integrated Bluetooth TWS earbud implementation. Some commercially available ANC earbuds are designed with small microphone capsules installed at the tips of their earplugs to serve as Feedback microphones for ANC functionality. By repurposing these existing microphones for speech pickup, we can quickly transform them into Whisphones. In some TWS earbuds, this internal microphone is utilized for voice pickup [7] during phone calls. However, the primary means of audio capture remains beamforming with multiple external microphones; the internal mic and vibration sensors inside the housing play auxiliary roles to enhance speech intelligibility. Therefore, we can convert these ANC earbuds into Whisphones by simply reprogramming the firmware to prioritize these feedback microphones as primary audio pickup channels. Here, we implemented this concept using PineBuds Pro [14], whose firmware is openly available online. Although its ANC performance is approximately 10dB or lower according to our measurements, leading to slightly higher WER values in noisy environments (as indicated by the dotted lines in Figure 3), it still achieves reasonably good WERs up to external noise levels of about 70dB(A).

## 4 Discussion

- Echo Cancellation

When Whisphone is used as integrated earphones, music or speech played through them can be picked up by the earplug-tip microphone and become a source of interference for voice recognition tasks and cause echoes during real-time calls. However, since the signals played back from the earbuds are known information, conventional echo cancellation techniques commonly employed in speakerphone applications can be used to address this issue effectively.

A simple approach involves using one side of stereo earbuds as a microphone and the other as an earphone when making calls or engaging in voice recognition tasks. This method has been implemented for the TWS-type device shown in Figure 4(c).

- Double Noise Reduction

In environments with background speech, certain words may still interfere with voice recognizers even after applying noise reduction techniques. To prevent unintended activation, one possible approach is to compare voice signals between two sources: outside (= ANC's FeedForward microphone) and inside (= ANC's Feedback microphone). By allowing only the sound components with higher volumes from the inside microphone to pass through, so that we can effectively block external speech interference. This method leverages

Whisphone: Whispering Input Earbuds

the fact that external noise will always be louder as captured by the outside microphone due to the physical blockage of the ear canals by earplugs.

- Monaural / Binaural

Whisphone offers flexibility in terms of wearing options, as it can be worn on either one ear (monaural) or both ears (binaural). In our experiments with a single-sided setup, we found no significant differences in recorded speech or noise levels when changing the conditions of the other ear canal, such as keeping it open, blocked by an earplug, or utilizing ANC functionality. Monaural use reduces the burden of wearing earbuds and allows for external sound perception while still providing adequate performance unless stereo music listening is intended. Binaural usage provides better articulation control over whispering through bone conduction auditory feedback, especially in high-noise environments.

Combining the outputs of microphones on both sides may further improve S/N ratios; theoretically, the maximum improvement is 1.4 times. However, this potential gain might be limited due to challenges in timing adjustments for TWS earbuds and decreased correlation between left and right speech components (as the ear canal and skull structures are not perfectly symmetrical), and also increased correlation among noise components (caused by bone conduction mixing).

- Improving Recognition Rates

Whispered speech captured via bone conduction through Whisphone has a significant reduction in high-frequency components above 2kHz. Certain phonemes rich in higher frequencies (e.g., the consonants /s/ and /h/, or the vowel /i/) may sound different from regular whispers, potentially leading to decreased recognition rates. Retraining speech recognition mechanisms using bone-conducted whisper data is the most effective solution; however, gathering extensive relevant voice samples may not always be feasible. As an alternative approach, if existing datasets for regular whispered speech are available, they can be filtered using a low-pass filter with a 2kHz cut-off frequency to simulate bone-conducted data. In addition, voiced sounds become unvoiced in whispered speech (e.g., /g/→/k/), but many mainstream voice recognition platforms have addressed this challenge by employing massive learning or whisper-specific word

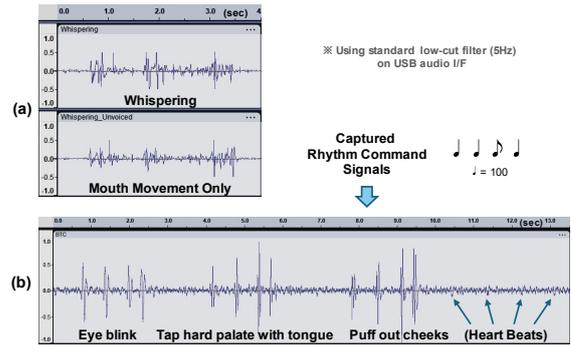

Figure 5. Exploring applications beyond speech signals: (a) Low-frequency components contain information about mouth movements during articulation, suggesting future possibilities for silent speech recognition by including these patterns in the training process; (b) Rhythmic blinking or tongue gestures can serve as hands-free and inconspicuous control commands.

dictionaries to mitigate any degradation in recognition rates effectively.

- Real-time Speech Conversion

While this paper primarily focuses on the recognition of bone conduction whispered speech for text generation applications, real-time conversion of captured whisper input into regular whispers or even normal speech enables additional use cases such as voice calls. For converting bone conduction whisper input to regular whispers, it is necessary to restore high frequencies above 2kHz; techniques for transforming bone conduction normal speech signals into air conduction signals can be employed [26]. To further convert whisper input into normal speech, restoration of vocal cord vibration components and pitch information becomes crucial; existing techniques for modifying whispered utterances into normal speech can be leveraged [17].

- Utilizing External Ear Canal Deformation during Utterance

Figure 5(a) illustrates sample waveforms obtained by Whisphone while whispering the phrase "Konnichiwa, watashi no koe ga kikoe masuka?" ("Hello, can you hear my voice?" in Japanese). The top part shows whisper input, and the bottom part depicts silent articulation of the same phrase without producing any sound (referred to as "silent speech"). Both waveforms exhibit similar spike-like low-frequency signals, which are believed to be pressure changes due to external ear canal deformation caused by mouth and jaw movements during speech production. Currently, these components are filtered out because their amplitudes exceed those of the target signal, leading to clipping distortion. However, as this component may contain some speech information, including it during the training of recognizers might further improve recognition performance. Furthermore, utilizing *only* these sub-audible components could enable silent speech recognition, offering enhanced privacy.

Moreover, Figure 5(b) shows waveform variations when performing rhythmic movements like blinking eyes, indicating that pressure changes in the external ear canal also contain motion information. This phenomenon opens up possibilities for hands-free operation through simple commands, especially those using tongue movements have a high level of confidentiality. Command functions using simple rhythms can be realized by using comparators and timers [27], making it suitable for low-power and always-on wakeup operations while offering better privacy protection compared to image or voice recognition methods. We have implemented this concept as a trigger command for AI assistant activation using our TWS earbud implementation shown in Figure 4(c). Furthermore, subtle heart rate information present within these pressure changes (see Figure 5(b) rightmost part) enables Whisphone to serve as a rudimentary heart rate sensor.

• Whispering Techniques

While regular speech production primarily originates from vocal cord vibrations, whispered utterances are generated by turbulent airflow passing through small gaps in the vocal tract. The methods for creating these gaps vary depending on individuals and situations; common techniques include narrowing the glottis (which also used for producing a husky voice), placing the tongue closer to the hard palate, or narrowing the lips. Narrowing the glottis creates more stable turbulence, allowing for pronouncing almost all voiceless consonants. However, this approach leads to reduced high-frequency components and decreased clarity while placing greater strain on the vocal cords. Whispering through lip movements makes it easier to decrease volume but restricts producible phonemes, especially nasal sounds like /n/. Using the palate offers characteristics between these two methods. To enhance Whisphone's recognition rates further, additional learning by sample data with various whispering technique might be necessary.

Additionally, we can whispering with ingressive speech (utterance while inhaling, described in SilentVoice [11]). Since the volume of ingressive whispers remains similar to regular (exhalatory) ones, it does not provide additional secrecy benefits over exhaled whispering. However, if Whisphone can distinguish between inhalation and exhalation whispering modes, perhaps by analyzing sub-audible external ear canal deformation components mentioned earlier, this capability could enable automatic switching for human interactions using whispered exhalatory speech versus AI assistant control via inhaled whisperings.

• Auto Ventilation

When wearing highly airtight canal-type earbuds during daily life, maintaining situational awareness and communicating safely with others require appropriate external sound intake. Most devices currently employ an "aware mode," where a portion of the external sounds captured by their outer microphones is played back through the earphones. However, this approach increases power consumption. Additionally, when wearing well-sealed canal earbuds, issues such as overly loud auditory feedback during speech and the perception of heartbeat can be distracting. Typically, these problems are addressed by providing vent holes under trading off noise suppression performance. When considering Whisphone for full-time wearability, addressing these challenges becomes crucial. One possible solution is to employ a (Normally-Open type) active ventilation mechanism using piezoelectric elements [23], allowing external "natural" sounds to pass through when power is off and electrically closing the vents during operation to block noise actively.

• Support for Other Languages

The Google Voice Search and OpenAI's Whisper that we used for evaluation are compatible with whispered speech recognition in multiple languages. We conducted similar experiments as described in Chapter 2 using English phrases and measured their WERs under silent conditions. While regular microphone-based whisper input yielded WER values of 2.4% and 0.0%, respectively, Whisphone's performance for English whispers deteriorated to 55.0% and 56.2%, respectively. English is a language dominated by high-frequency consonants above 1kHz; thus, it suffers from the reduced high-frequency components in bone conduction whispering. In addition, Mandarin Chinese, being a tonal language, may face challenges with Whisphone due to the lack of tonal information in whispered speech. In this regard, Whisphone is more suitable for "CV" languages like Japanese and Korean, which rely primarily on vowels with minimal tone changes.

## 5  Conclusion

In this paper, we introduce Whisphone – a full-time wearable whisper input device designed to transform computers into our conversational partners in daily life. We initially target the adoption of Whisphone in Japanese society, which highly values both consideration for others and linguistic advantages. With ongoing improvements to overcome language barriers, we envision the widespread adoption of Whisphone globally. While this paper primarily focuses on speech input applications, real-time (both online and offline) conversations using Whisphone also offer the potential advantage of minimizing disturbances to those nearby and preventing unintended information leaks.

The quiet future is coming.

Whisphone: Whispering Input Earbuds

---

Future Vision: Evolution or Regression?

Humankind has primarily relied on speech for person-to-person communication since ancient times. As we continue to coexist with cyberspace, an instantaneous and reliable means of switching between human and network conversations will become essential (wake words or gestures cannot easily eliminate the risk of accidental input errors). However, finding a new mode that doesn't interfere with our muscles – our only actuators, already occupied with daily tasks – is challenging. Utilizing vestigial organs like ear movement muscles [24] may require some training but holds promise as an essential skill for future generations to seamlessly interact with AI: imagine lifting your ears slightly while whispering for discrete voice input commands (perhaps we need to start practicing with morning exercises of a hundred ear lifts!).